\documentclass[11pt,a4paper]{article}
\usepackage{arxiv}

\usepackage[utf8]{inputenc} % allow utf-8 input
\usepackage[T1]{fontenc}    % use 8-bit T1 fonts
\usepackage{hyperref}       % hyperlinks
\usepackage{url}            % simple URL typesetting
\usepackage{booktabs}       % professional-quality tables
\usepackage{amsfonts}       % blackboard math symbols
\usepackage{nicefrac}       % compact symbols for 1/2, etc.
\usepackage{microtype}      % microtypography
\usepackage{lipsum}
\usepackage{bm,amssymb,amsmath,float,caption,multicol,graphicx}
\usepackage[justification=centering]{subcaption}

\title{\texttt{rstap}: An R package for Spatial Temporal Aggregated Predictor Models}

\author{
  Adam Peterson \\
  Department of Biostatistics\\
  University of Michigan\\
  Ann Arbor, MI 48104 \\
  \texttt{atpvyc@umich.edu} \\
   \And
 Brisa S\'{a}nchez \\
  Department of Biostatistics\\
  University of Michigan\\
  Ann Arbor, MI \\
  \texttt{brisa@umich.edu} \\
}

\date{}

\begin{document}
	
	\maketitle
	\begin{abstract}
    The \textbf{rstap} package implements Bayesian spatial temporal aggregated predictor models in R using the probabilistic programming language \texttt{Stan}.  A variety of distributions and link functions are supported, allowing users to fit this extension to the generalized linear model with both independent and correlated outcomes. 
	\end{abstract}
    
	\section{Introduction}
	
     The built environment refers to the human made space in which people live, work, and recreate on a day-to-day basis \cite{roof2008public}. The features of the built environment are many - ranging from sidewalk availability, street density, green space, ambient light or sound levels to the physical presence of amenities like community centers or businesses that can be mapped as point locations according to their address.  In this paper we are primarily concerned with the latter and for simplicity refer to them as built environment features (BEFs). An expanding body of research is focused on quantifying the health impact of BEFs since they constrain and enable everyday choices that may contribute to the development of disease \cite{booth2005obesity,charreire2010measuring}. For example, in the United States, a preponderance of convenience stores or fast food restaurants surrounding schools offering predominantly ``junk''  food may increase the odds an attending student develops obesity
   \cite{davis2009proximity,sanchez2012differential,currie2010effect,harris2011location,langellier2012food}.\par
    A key limitation in this research area is that the spatial and temporal scales at which BEFs may impact health are unknown; that is, there is uncertainty about the domain, in either space or time, at which BEFs are most relevant for health. One method in current use that has focused on addressing questions about the spatial scale is the distributed lag model (DLM) \cite{baek2016distributed,baek2016hierarchical}. DLMs use counts of businesses within a discrete series of user-specified distance radii to examine how effects of BEFs decay as a function of distance between BEFs and study participants. However, DLMs are limited by the fact that they do not specifically estimate a spatial scale parameter and require a sufficient number of distance radii parameters \cite{baek2016distributed,gasparrini2014modeling,guo2011impact}. \par
      Peterson et al \cite{petersonsanchez} recently proposed spatial-temporal aggregated predictor (STAP) models to address questions about both the spatial-temporal scales and to avoid the accuracy issues associated with discretizing distances in the DLM.  The purpose of the present article is to provide a general overview of the \texttt{rstap} package (version 1.0.2), which implements STAP models. This package utilizes Hamiltonian Markov Chain Monte Carlo (HMC) estimation via \texttt{stan}\cite{carpenter2016stan}. \texttt{rstap} can be considered as being in the same family of packages as \texttt{brms} \cite{brms} and \texttt{rstanarm} \cite{rstanarm} in the sense that it utilizes \texttt{Stan} as the estimation engine to fit models of a designated form. We begin with a review of the modeling framework followed by an introduction of the software using both simulated and real data.  Throughout we use food outlets as examples of BEFs - e.g. fast food restaurants and coffee shops - and body mass index (BMI) or obesity as examples of outcomes. We end by discussing current limitations and future plans for extending the package and modeling framework.
	\section{Model Description}
	Let $\bm{Y}_i$ be a vector containing $n_i$ repeated outcome measures for subject $i= 1, 2, ..., N$, with corresponding mean vector $\bm{\mu}_i = [\mu_1,...,\mu_{n_i}]$. The modeling objective is to estimate a generalized linear mixed model (GLMM) regression of $\bm{Y}_i$ on typical covariates and summaries of built environment features. Define $\bm{Z}_i$ as a matrix of covariates with corresponding population parameters $\bm{\delta}$ for said regression and let $\bm{b}_i \sim MVN(\bm{0},\Sigma)$ be random effects for subject $i$, with corresponding design matrix $\bm{W}_i$. The STAP model extends the standard GLMM, $g(\bm{\mu}_i) = \alpha + \bm{Z}_i \bm{\delta} + \bm{W}_i \bm{b}_i$, where $g(\cdot)$ is a link function, by estimating the latent effect of BEFs on $\bm{\mu}_i$. For simplicity, we describe the model focusing on one type of BEF, fast food restaurants (FFR), and univariate outcome $\mu_i \in \mathbb{R}^1$, discussing how it can be expanded later. \par
	In order to estimate the BEF effect and spatial scale, the STAP model requires pairwise distances $d$ between each subject and each BEF, in addition to a weighting function  $\mathcal{K}_s$ chosen so that $K_s(0,\theta) = 1, \lim_{d \to \infty} \mathcal{K}_s(d,\theta) = 0$ where distance $d \in [0,\infty)$ and spatial scale $\theta \in \mathbb{R}^{+}$. This corresponds to the substantive belief that a given BEF's maximum impact is made when a subject is as close as possible to it and vice versus. \par
	Let $\mathcal{D}_{i}$ be the set of all aforementioned pairwise distances, $d$, between all the BEF locations and subject $i$. The model is then:
    \begin{align*}
    g(\mu_i) &= \alpha + \beta X_{i}(\theta) + \bm{Z}_i^{T}\bm{\delta}   \tag{1}\\
    X_{i}(\theta) &= \sum_{d \in \mathcal{D}_{i}} \mathcal{K}_s \big( \frac{d}{\theta^{s}} \big )
    \end{align*}
    Hence, $X_{i}$ represents subject $i$'s  cumulative exposure to the particular type of BEF, accumulated according to the spatial scale $\theta^{s}$. The coefficient $\beta$ is the estimated difference in $g(\bm{Y}_i)$ associated with a one unit increase in $X_{i}(\theta)$ all else equal. Equivalently, $\beta$ can be thought of as the change in $g(\bm{Y}_i)$ associated with placing one new BEF at a distance of $0$ units from the subject, all else equal, since this placement results in one unit higher $X_{i}(\theta)$. A larger spatial scale corresponds to a BEF that continues to have an impact at larger distances and vice versus. \par 
    
    One may consider having multiple types or classes of BEFs in the same model. For instance, consider FFRs in addition to convenience stores and gyms. The model is extended by indexing the classes of BEFs by $j=1, \dots, J$, and re-writing  model (1) as: 
     \begin{align*}
    g(\mu_{i|\bm{b}_i}) &= \alpha + \sum_{j=1}^J \beta_j X_{i,j}(\theta_j)+ \bm{Z}_i^{T}\bm{\delta}  + \bm{W}_i^{T}\bm{b}_i \tag{2}\\
    X_{i,j}(\theta_j) &= \sum_{d \in \mathcal{D}_{ij}} \mathcal{K}_s \big( \frac{d}{\theta^{s}_j} \big )
     \end{align*}
      where $\mathcal{D}_{ij}$ is the set of distances from individual $i$ to BEFs of class $j$ associated with subject $i$, $\theta_j$ is a spatial scale specific to the $j$th class and $g(\bm{\mu}_{i|\bm{b}_i})$ is the transformed mean of outcome $\bm{Y}_i$ conditional on $\bm{b}_i$, $g(E[\bm{Y}|\bm{b}_i])$. \par
      Extending this framework to a setting in which $\bm{\mu}_i \in \mathbb{R}^{n_i}$ and defining a model with both temporal and spatial components under the STAP framework yields the following expression:
    \begin{align*}
    g(\bm{\mu}_{i|\bm{b}_i}) &=  \alpha + \sum_{j=1}^J \beta_j \bm{X}_{i,j}(\bm{\theta})  +\bm{Z}_i\bm{\delta} + \bm{W}_i\bm{b}_i \tag{3}\\
    X_{i,j,k}(\bm{\theta}) &= \sum_{(d,t) \in \mathcal{D}_{ijk}} \mathcal{K}_s \big( \frac{d}{\theta^{s}_j} \big )\mathcal{K}_t(\frac{t}{\theta^{t}_j})
    \end{align*}
    Here $t \in [0,\infty)$ represents the time subject $i$ has spent ``exposed'' to BEFs of class $j$ at distance $d$ by the $k$th measurement visit $(k = 1,..., n_i$). Thus, $X_{i,j,k}(\bm{\theta})$ represents the exposure of the $i$th subject accumulated by the $k$th visit attributable to all BEFs in the $j$th class assuming the spatial and temporal components of the BEF exposure are independent.  The scales for BEF of class $j$, are $\bm{\theta}_j = (\theta_j^s,\theta_j^t)$ where $\theta_j^{t}$ represents the temporal scale at which exposure to a given BEF is associated with $g(\bm{\mu}_i)$.  In contrast to $\mathcal{K}_s$, the weight function $\mathcal{K}_t$ is chosen so that $\mathcal{K}_t(0)=0,\lim_{t\to \infty} \mathcal{K}_t(t)=1$; $\beta_{j}$ in this setting is interpreted as the difference in $g(\bm{\mu}_i)$ when a BEF is placed at distance 0 from the subject, for an amount of time that approaches infinity. Note here that the statement regarding infinite time will always be true but the BEF may exert 99$\%$ of its impact in a finite amount of time when the temporal scale $\theta_j^t$ is small. 
    
    \section{Parameter Estimation}
    
    All the above models are fit according to a Bayesian framework in the \texttt{rstap} package, allowing the incorporation of prior information into the model and the final goal being posterior inference. To that end, the \texttt{rstap} package utilizes \texttt{stan} to fit models, drawing samples from the posterior distribution of model parameters obtained from an extension of the static HMC sampler \cite{neal2011mcmc}, the Generalized No-U Turn Sampler (NUTS) \cite{hoffman2014no,betancourt2017conceptual}. ``Hamiltonian Monte Carlo (HMC) is a Markov chain Monte Carlo (MCMC) algorithm that
avoids the random walk behavior and sensitivity to correlated parameters that plague many
MCMC methods by taking a series of steps informed by first-order gradient information.
These features allow it to converge to high-dimensional target distributions much more
quickly than simpler methods such as random walk Metropolis\cite{metropolis1953equation,hastings1970monte} or Gibbs sampling'' \cite{hoffman2014no}. This trade off comes at the expense of calculating the gradient of parameters to be estimated and subsequent numerical integration to solve the Hamiltonian differential equations \cite{carpenter2016stan,betancourt2017conceptual,gelman2013bayesian}. \par
     Consequently this framework both enables and constrains the kinds of models that can currently be fit in \texttt{rstap}. Since each $X_{i,j}$ is a function of $\theta_j$, several distances and/or times, the time required to differentiate $\theta_j$ and $\beta_{j}$ via \texttt{Stan}'s automatic reverse mode differentiation scheme is longer than it would be in a typical regression setting. The computational complexity of this differentiation is a function of both the specification of $\mathcal{K}_{s}$ and/or $\mathcal{K}_t$, as well as the number of BEFs.  To provide a bound on this complexity in \texttt{rstap}, the number of BEFs is constrained to those that lie within a pre-specified maximum distance. This maximum distance also enforces an upper bound on the possible values of $\theta^{s}$. Since the monotonic function $\mathcal{K}_s$ discussed previously would become uniform for an infinitely large scale $\theta^s$ evaluated at finite distance, an upper bound must be chosen to ensure the posterior of the scales do not tend toward infinity in the case where the estimated spatial exposure approaches uniform on the domain of distances under which the model was fitted. The current setting in \texttt{rstap} divides the max distance by the maximum ninety seventh quantile of the chosen $\mathcal{K}$ function to obtain this upper bound - an explicit formulation of this bound can be found in the Appendix, Figure \ref{fig:theta_up}. This allows users to recognize the ``approximate'' uniformity of the exposure effect of a BEF and re-adjust the inclusion distance as needed. An example of this is given on the \texttt{rstap} website (https://biostatistics4socialimpact.github.io/rstap/articles/Introduction.html).\par
     With our simulated and real datasets examining between one and five STAPs with anywhere from a few hundred to a few thousand subjects, and anywhere from a few dozen to a few hundred pairwise distances per person,  sampling times for 500 posterior samples - including warmup - vary from a few seconds to just under an hour. Further comments on these constraints and future work for plans to speed these sampling times are elaborated upon in the discussion. \par
    Similar to \texttt{rstanarm} and \texttt{brms}, \texttt{rstap} also offers draws from the posterior predictive distribution and pointwise log-likelihood. The former allows assessment of model fit between the model's outcome prediction's, $\hat{\bm{Y}}$, and $\bm{Y}$, while the latter permits model selection via Watanabe-Akaike information criteria \cite{vehtari2017practical}. Software functions in \texttt{rstap} that take advantage of these features are discussed in our example analyses.
    \section{Software}
     The current development version of \texttt{rstap} can be downloaded and installed from GitHub via  \texttt{devtools::install\_github("biostatistics4socialimpact/rstap")}. \par
     Models are fit in \texttt{rstap} using the following procedure, which is also summarized in Figure 1. This framework is akin to the \texttt{lm} and \texttt{glm} functions in the popular \texttt{stats} package \cite{Rlanguage} as well as the \texttt{lmer} function in \texttt{lme4} for correlated data \cite{bates}. A user provides a formula which specifies the kind of model, the family of distributions and the link function under which the mean of the distribution is hypothesized to be related to the linear predictors. Currently Binomial, Bernoulli, Poisson and Gaussian distribution families are supported by \texttt{rstap}.\par
     In contrast to \texttt{lme4} and \texttt{stats}, a total of two or three data sets must be supplied to the function call, each containing an ID to relate them to each other.  One of these datasets, ``subject\_data'', must always be supplied to the function with a unique subject identifier on each row, any potential group IDs, and any other standard covariates to include in the model.  The other two datasets  contain the built environment information: specifically, the distances or times between subjects and BEFs. One or both of these are required dependent upon whether only or both spatial and temporal components are specified in the model. The typical structure for built environment datasets - ``distance\_data'' and ``time\_data'' - is given below in Table \ref{tab:distdata1}.  Each row defines a unique subject BEF association with its corresponding distance. The \texttt{subject\_ID} argument in this model would be the, identically named, ``subject\_ID'' string, since that is the name of the column containing the unique subject ID that maps subjects' distances in the ``distance\_data'' data frame to their appropriate  covariate information in the ``subject\_data'' data frame. In the case of grouped data, an additional column would define the group to which the data should be associated - see Tables \ref{tab:distdata2} and \ref{tab:timedata} for examples.\par
     These data are then passed to one of several pre-set compiled \texttt{Stan} programs akin to \texttt{rstanarm} for sampling. After this a \texttt{stapreg} object will be defined using the samples. The \texttt{stapreg} class is modeled off of the \texttt{stanreg} class in \texttt{rstanarm} allowing a user to print out a clean summary of the model estimates including diagnostics of the MCMC fit, such as MCMC standard errors and standard split-chain $\hat{R}$ diagnostics \cite{gelman2013bayesian}.
\begin{table}[H]
    \centering
\begin{tabular}{c|c|c|c|c}
subject\_ID  & bef\_ID & bef\_name    & Distance \\ \hline
1           & 1       & Fast\_Food   & 0.351    \\
1           & 2       & Fast\_Food   & 0.891    \\
2           & 1       & Fast\_Food   & 1.231    \\
2           & 2       & Fast\_Food   & 0.331    \\
2           & 3       & Coffee\_Shop & .531    
\end{tabular}
\caption{Example data structure for distance data}
\label{tab:distdata1}
\end{table}
    \subsection{A Worked Example}
    There are two main functions in the \texttt{rstap} package: \texttt{stap\_glm} and \texttt{stap\_glmer}. Two other functions included in the package, \texttt{stap\_lm} and \texttt{stap\_lmer} have the same purpose as the previous, but simply set the family option to be \texttt{gaussian()}, implying a gaussian distribution with an identity link will be used to fit the data. \par
   Suppose one has a subject level dataset with subjects' sex and BMI recorded, as well as a dataset with the distances, in miles, between each subject and all FFRs within an area of, say, 5 miles of each subject. Then the code in Figure \ref{fig:stapglm1} would correspond to the following model and prior distribution specifications, relating the average BMI to subjects' sex and exposure to FFRs:
    \begin{align*}
    \text{BMI}_i &= \alpha + \delta\text{Sex}_i + \beta\text{Fast\_Food}_i(\theta^{s}) + \epsilon_i\\
    \epsilon_i &\stackrel{iid}{\sim} N(0,\sigma) \quad \quad \sigma \sim C^{+}(0,5)\\
    \beta & \sim N(0,4) \quad \quad \delta  \sim N(0,4)\\
    \alpha &\sim N(26,4) \quad \quad \log(\theta^s) \sim  N(1,1)
    \end{align*}
    Further details regarding the specification of the priors are discussed in Section 4.3.
    
    \begin{figure}[H]
    \begin{verbatim}
    R> fit <- stap_glm(formula = BMI ~ sex + sap(Fast_Food),
                       family = gaussian(link="identity"),
                       subject_data = subject_data,
                       distance_data = distance_data,
                       subject_ID = "subject_ID",
                       max_distance = 5,
                       ## prior for delta
                       prior = normal(location = 0, scale = 4, autoscale = F),
                       ## prior for alpha - possibly standardized
                       prior_intercept = normal(location = 26, scale = 4, autoscale = F),
                       ## prior for beta - always standardized
                       prior_stap = normal(location = 0, scale = 4), 
                       ## not standardized
                       prior_theta = log_normal(location = 1, scale = 1), 
                       ## folded cauchy
                       prior_aux = cauchy(location = 0, scale = 5, autoscale = F), 
                       iter = 2E3, warmup = 1E3, chains = 4, cores = 4)
    \end{verbatim}
    \caption{Typical Syntax for fitting a stap model via rstap}
    \label{fig:stapglm1}
    \end{figure}
    \vspace{-2.cm}
    \subsection{Formula}
    
     The formula argument to all \texttt{rstap} functions contains the relationship between the response and the predictors, as well as the specific configuration for the spatial-temporal predictors or group terms included in the model. Group terms are specified identically to \texttt{lme4}'s syntax, where \texttt{(coefs|group)} denotes that a random intercept and slope for ``coefs'' at the level of ``group'' should be included in the model. Spatial-Temporal components may be specified in a number of ways: Table \ref{tab:frmla} below shows the different syntax that may be used to fit a model akin to the one in Section 4.1, with differing spatial-temporal components using the two different weight functions implemented in \texttt{rstap}. Note that while \texttt{sap}, \texttt{tap} and \texttt{stap} each refer to different kinds of STAPs, references made to \texttt{stap}s hereafter encompass all three kinds of covariates unless stated otherwise.  

\begin{table}[H]
\begin{tabular}{l|l|l|l}
\hline
\multicolumn{4}{|l|}{Default setting configures $\mathcal{K}$ to the error (temporal) or complimentary error function (spatial)}                                                 \\ \hline
Formula                                            & Kind of STAP predictor & $\mathcal{K}_s(d,\theta^{s})$        & $\mathcal{K}_t(t,\theta^{t})$    \\ \hline

BMI $\tilde{}$ sex + sap(Coffee\_Shop)             & Spatial                & $ 1- \text{erf}(\frac{d}{\theta^s})$ & -                                \\
BMI $\tilde{}$ sex + tap(Coffee\_Shop)             & Temporal               & -                                    & erf$(\frac{t}{\theta^t})$        \\
BMI $\tilde{}$ sex + stap(Coffee\_Shop)            & Spatial - Temporal     & $1$- erf$(\frac{d}{\theta^s})$       & erf$(\frac{t}{\theta^t})$        \\
BMI $\tilde{}$ sex + sap(Coffee\_Shop, exp)        & Spatial                & exp$(-\frac{d}{\theta^{t}})$         & -                                \\
BMI $\tilde{}$ sex + tap(Coffee\_Shop, cexp)       & Temporal               & -                                    & 1 - exp$(-\frac{t}{\theta^{t}})$ \\
BMI $\tilde{}$ sex + stap(Coffee\_Shop, exp, cexp) & Spatial - Temporal     & exp$(-\frac{d}{\theta^{t}})$         & 1 - exp$(-\frac{t}{\theta^{t}})$
\end{tabular}
\caption{Comparison of STAP model syntax across different weight functions predictor types. erf is the error function defined as erf$(x):=\frac{2}{\sqrt{\pi}} \int_{0}^{x} e^{-u^{2}}du$}
\label{tab:frmla}
\end{table}

    \subsection{Prior distributions}
    
    Discussion of how priors should be placed on parameters in standard regression models has already been elaborated in the packages mentioned previously \cite{brms,rstanarm}. In the next two sections we briefly discuss new considerations and \texttt{rstap} syntax for setting priors on the STAP specific variables: $\theta^s, \theta^t, \beta$.
    
    \subsubsection*{Spatial-Temporal Scale Parameters}
    \texttt{rstap} allows for custom specification of the priors on any scale parameters, $\theta^{s},\theta^{t}$ . This allows each scale to have its own prior distribution according to the user's desire. The syntax used in Figure \ref{fig:stapglm1}, corresponds to assigning the same prior distribution to all spatial-temporal scales, $\theta^s,\theta^t$. In cases where there is only one spatial predictor or there are similar \textit{a priori} beliefs about the spatial-temporal scales, this allows for easy assignment of scale priors. Alternatively, if one were interested in assigning different priors, perhaps when fitting a model with both spatial and temporal components on the class of ``Fast Food'' restaurants from the previous example, the model and corresponding code would look as follows:
    \begin{align*}
    \text{BMI}_i &= \alpha + \delta \text{Sex}_i  + \beta\text{Fast\_Food}_i(\theta^s,\theta^t)  + \epsilon_i\\
    \log(\theta^s) &\sim N(1,1) \quad \quad \log(\theta^t) \sim N(1,2)
    \end{align*}
    \begin{figure}[H]
    \begin{verbatim}
    R> fit <- stap_glm(formula = BMI ~ sex + stap(Fast_Food), ..., 
                       prior_theta = list(Fast_Food = 
                                     list(spatial = log_normal(location = 1,
                                                               scale = 1),
                                      	  temporal = log_normal(location = 1,
                                      	                        scale = 2))))
    \end{verbatim}
    \caption{Model syntax for explicitly specifying different priors for different spatial-temporal scales.  The ellipsis here indicates other arguments can be the same as specified in Figure 1}
    \label{fig:stapglm2}
    \end{figure}
    Currently only log and folded normal prior distributions are supported for $\theta^{s},\theta^{t}$ in \texttt{rstap} as these are the only priors with which we have done extensive simulations and testing. These can be used for less and more informative priors, respectively.
    
\subsubsection*{Population and group level parameters}
 Population and group level parameters are constructed and assigned priors in essentially the same form as they are in the \texttt{rstanarm} \cite{rstanarm} package. The only difference is that there are separate \texttt{prior} and \texttt{prior\_stap} arguments to allow for differing prior specifications for the $\bm{\delta}$ and $\bm{\beta}$ parameters, respectively. Note that while the user can choose to autoscale priors for the standard regression coefficients as in \texttt{rstanarm}, priors are always set on standardized scales for $\beta$ parameters. That is, during estimation, the estimated $X(\theta)$ values are mean centered and scaled to allow for easier computation, making the parameter space for a $\beta$'s far easier to traverse for the sampler \cite{carpenter2016stan,gelman2013bayesian}. Further, this allows placement of equivalent priors to be set on multiple \texttt{stap} covariates that might otherwise be inappropriate due to highly differing exposure levels \cite{gelman2013bayesian}. \\

    \section{Control}
    Regular control of the sampler will occur via three arguments: \texttt{iter}, \texttt{warm\_up}, and \texttt{adapt\_delta}. The first controls the total number of iterations for which the sampler is run. The second controls the subset of those previous iterations that are used to ``warm-up'' or tune the number of steps and mass-matrix which are used to propose new samples via NUTS\cite{hoffman2014no}. Finally, \texttt{adapt\_delta} controls the target sample acceptance ratio which determines the resolution at which the sampler explores the posterior. A higher \texttt{adapt\_delta} corresponds to a higher target ratio and consequently, a longer sampling time. Two additional arguments, \texttt{cores} and \texttt{chains}, specifies the number of processors to use and the number of chains to run. For example, if \texttt{chains = 2} and \texttt{cores = 2} then two chains will be run in parallel via the \texttt{parallel} package \cite{Rlanguage}. Further specification of \texttt{stan}'s algorithm specific variables occurs via the optional argument \texttt{control} analogous to \texttt{rstanarm} \cite{rstanarm}.
    
    \section{Example Analyses}
    In the following four sections, we demonstrate the STAP framework in settings involving continuous and binomial outcomes from both simulated and real data.

    \subsection*{Simulation I: Uncorrelated Outcomes}
\begin{figure}[H]
    \centering
    \begin{subfigure}{.45 \textwidth}
    \includegraphics[width = .85 \textwidth]{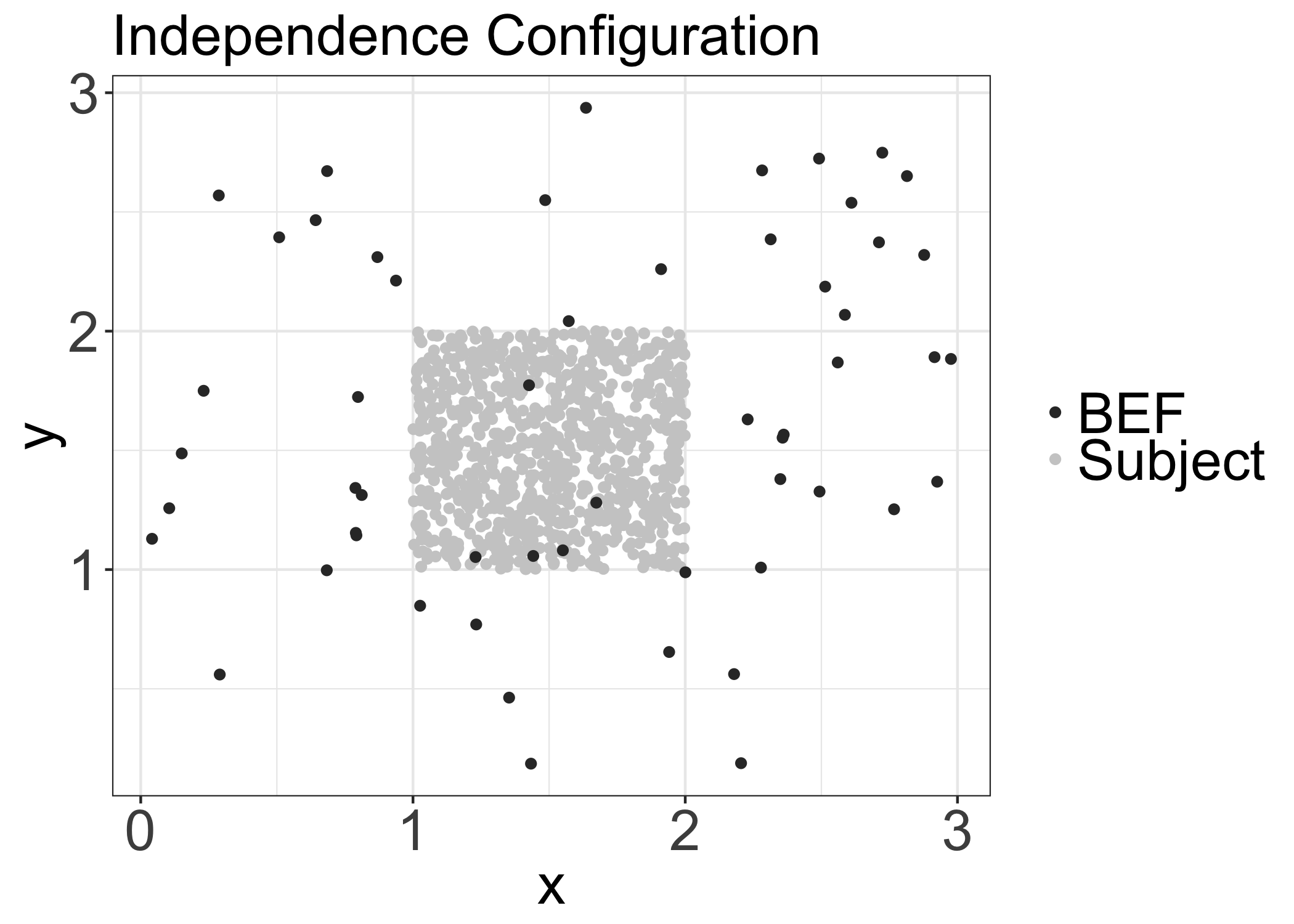}
    \caption{Spatial Arrangement of Subjects, Built Environment Features.}
    \end{subfigure}
    \begin{subfigure}{.45 \textwidth}
    \includegraphics[width = .85 \textwidth]{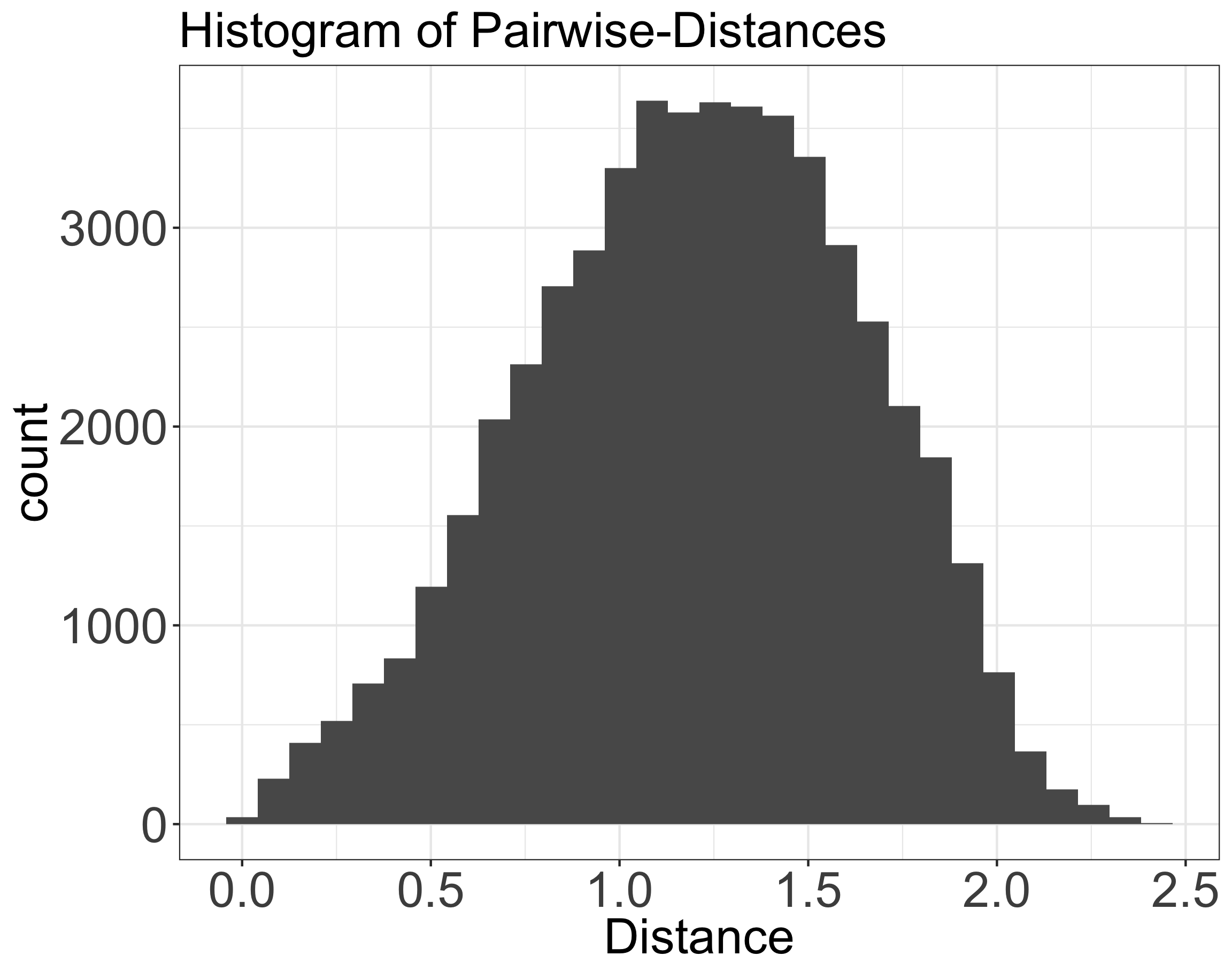}
    \caption{Histogram of Pairwise Euclidean Distance for simulated data.}
    \end{subfigure}
    \caption{Descriptive Graphics for Simulated Data}
    \label{fig:simIeda}
    \end{figure}

\begin{figure}[H]
    \centering
    \begin{subfigure}{.45 \textwidth}
    \includegraphics[width = .85 \textwidth]{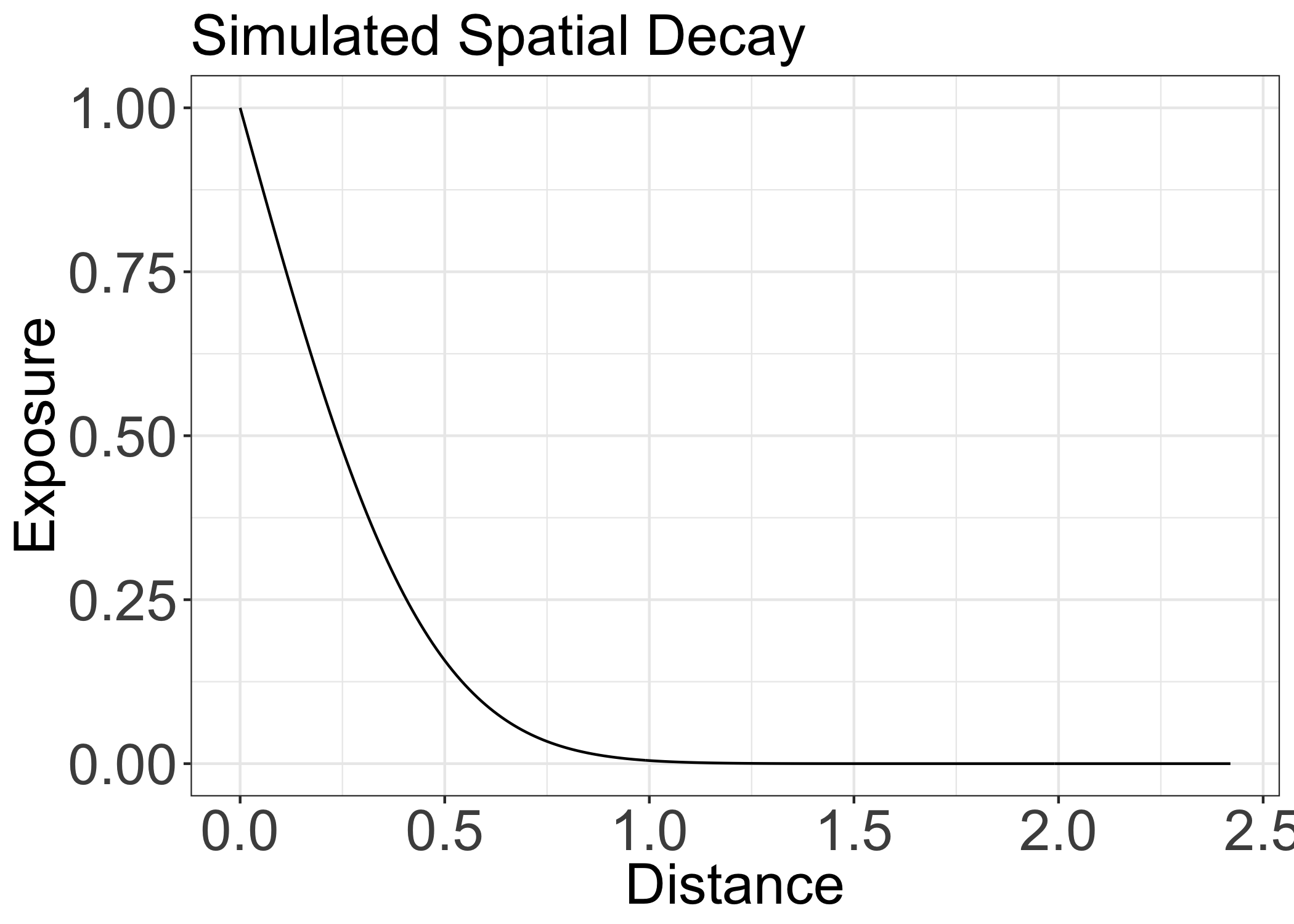}
    \caption{Simulated ``Fast Food'' Spatial Weight Function.}
    \end{subfigure}
    \begin{subfigure}{.45 \textwidth}
    \includegraphics[width = .85 \textwidth]{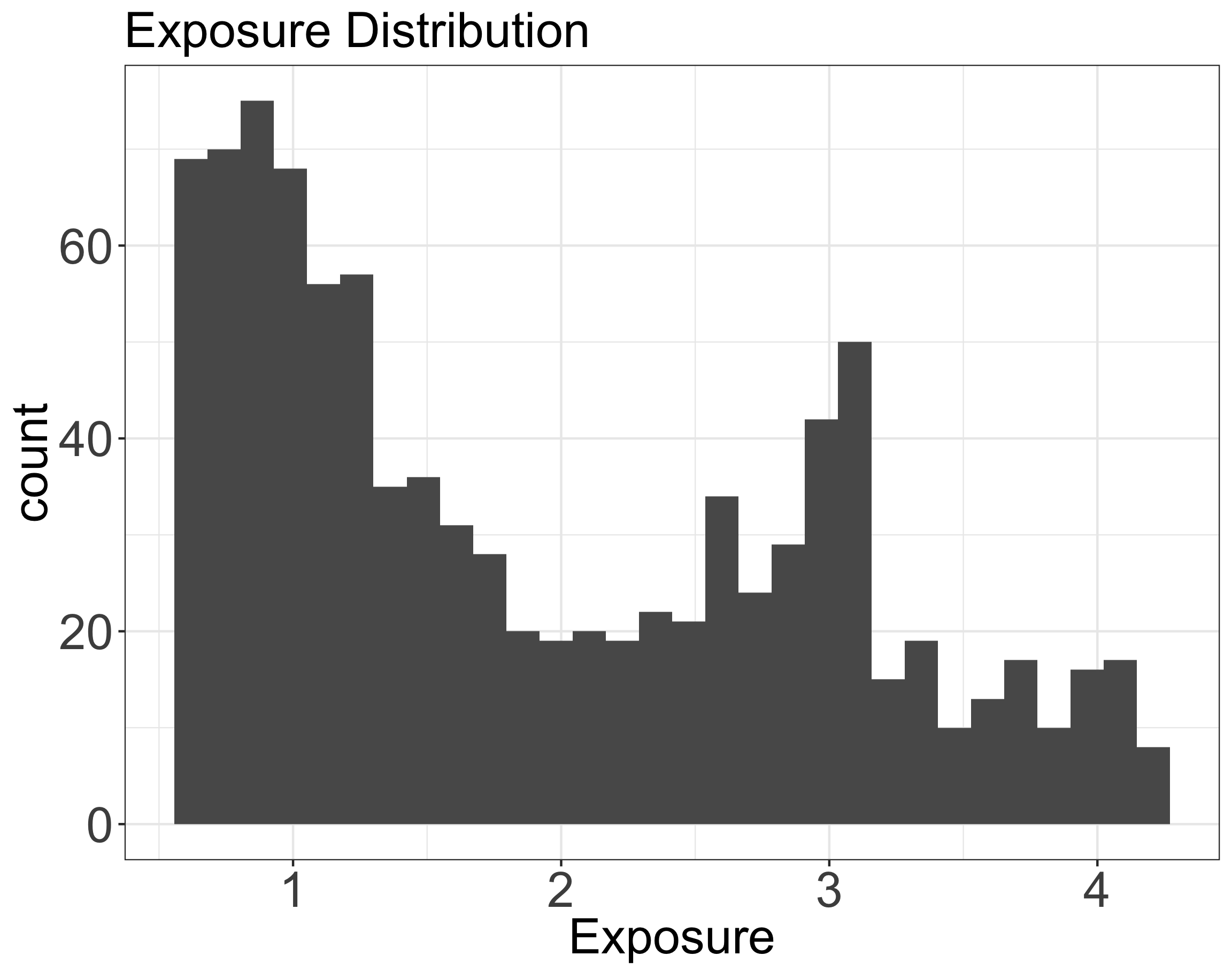}
    \caption{Histogram of ``Fast Food'' Exposure Distribution.}
    \end{subfigure}
    \caption{Spatial Aggregated Predictor Descriptive Graphics}
    \label{fig:simIerf}
    \end{figure}

    We demonstrate the \texttt{rstap} package first on a simulated dataset where the built environment features are homogeneously distributed in a 3 x 3 square, and the subjects are similarly homogeneously distributed in a 1 x 1 square centrally interior to the latter. Although obviously fictitious, this set-up provides some intuition as to how the different components of the model can be set-up and interpreted. The spatial arrangement and distribution of pairwise distances between subjects and BEFs can be seen in Figure \ref{fig:simIeda}. \\
    \indent 
    The outcome is then simulated according to the following model, with the intention of mimicking the marginal distribution of BMI in a random sample of subjects. Labeling our simulated BEFs as FFRs and simulating the spatial exposure using the complementary error function,  we recreate the example previously discussed in 4.1. Visualizations of the spatial decay and resulting exposure distribution can be seen in Figure \ref{fig:simIerf}.
    \begin{align*}
   BMI_i &= 22.5 - \text{Sex}_i 0.8 + \text{Fast\_Food}(\theta^{s}=.5) 1.2 + \epsilon_i\\
   \epsilon_i &\sim N(0,2.3)
    \end{align*}
    \vspace{-1cm}
    
    Assuming the data has been appropriately structured as described in Section 4, the model can then be fit with the syntax in Figure \ref{fig:stapglm1} after loading the \texttt{rstap} library. Typical model printout is seen below in Figure \ref{fig:stapglmprintout}.
    
    Calling \texttt{summary()} on the fit object will produce a longer print out, with relevant convergence diagnostics and WAIC, if specified - Appendix Figure \ref{fig:sumex}. Furthermore, typical model fit diagnostics such as posterior predictive checks \cite{gelman2013bayesian} can be accessed directly through the \texttt{rstap} \texttt{posterior\_predict} function and the \texttt{ppc\_dens\_overlay} function from the \texttt{bayesplot} package\cite{bayesplot}. The posterior predictive checks and NUTS energy diagnostics \cite{betancourt2017conceptual} - available again through the \texttt{bayesplot} package - for this model can be seen in Figure 5 and the corresponding code can be found in the Appendix - Figure \ref{fig:ppcs} \\
    
    \begin{figure}[H]
    \centering
    \begin{subfigure}{.45 \textwidth}
    \includegraphics[width = .85 \textwidth]{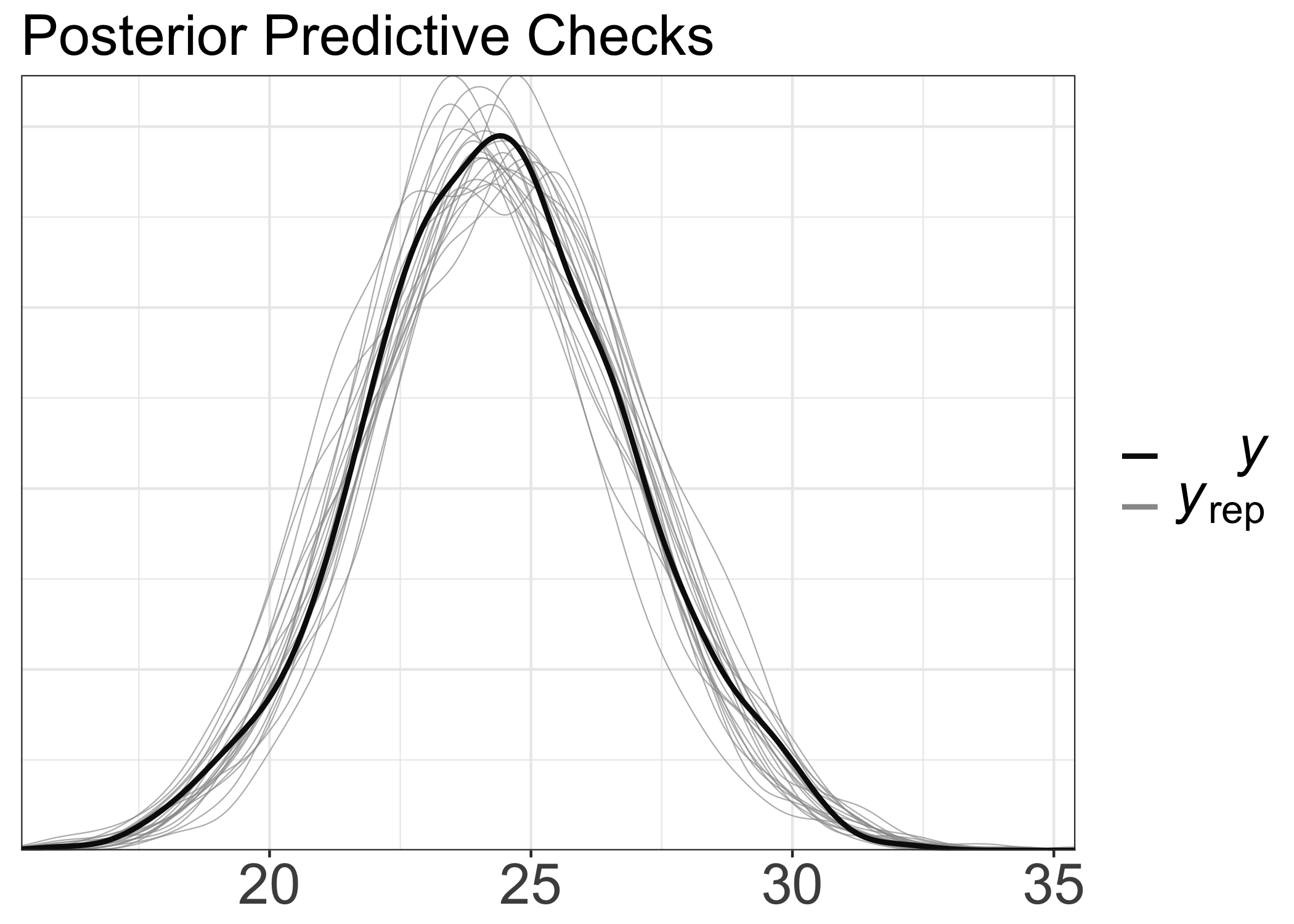}
    \end{subfigure}
    \begin{subfigure}{.45 \textwidth}
    \includegraphics[width = .85 \textwidth]{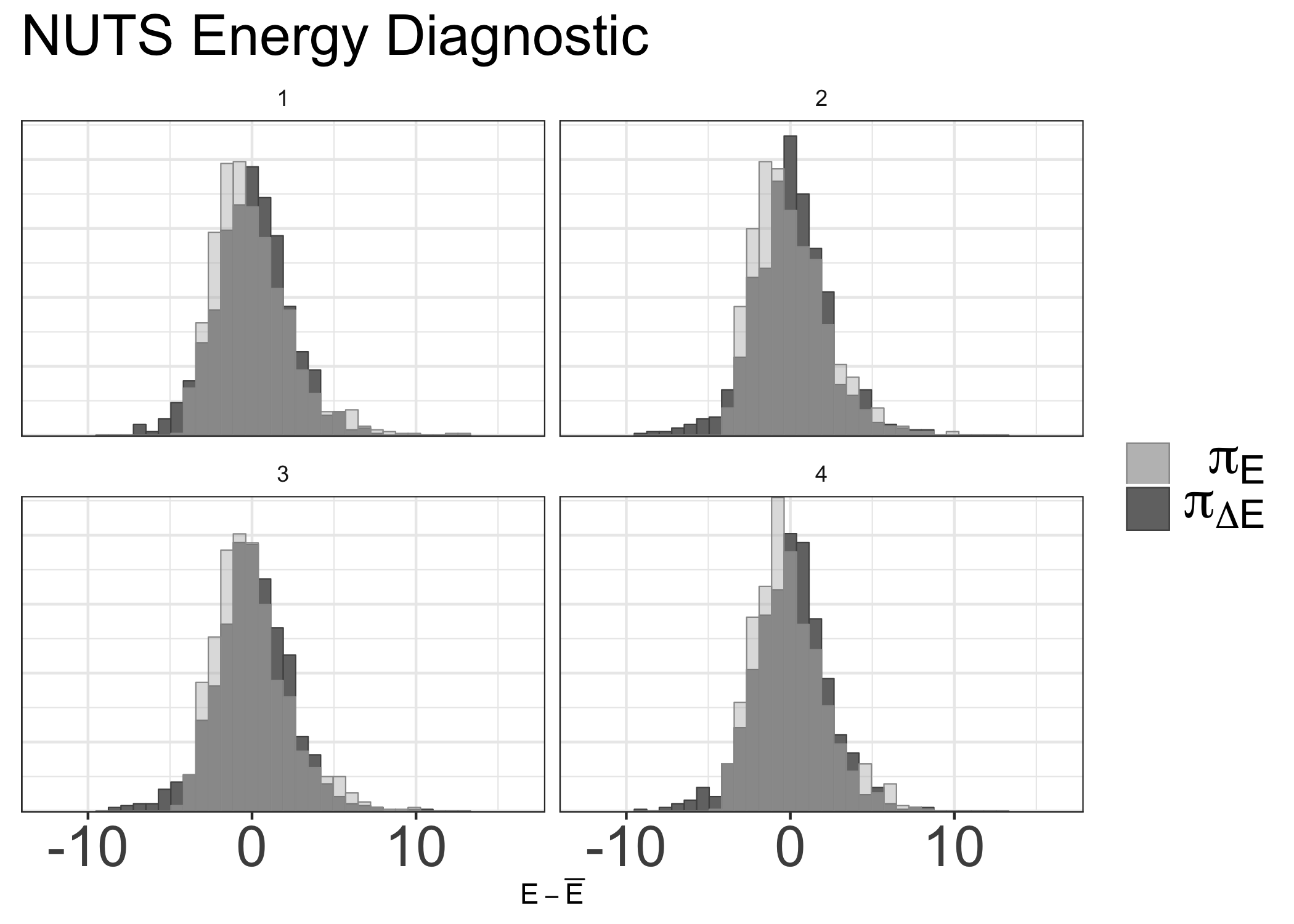}
    \end{subfigure}
    
    \caption{Posterior Predictive Checks (left), NUTS Energy Diagnostic (right) for Simulation I}
    \label{fig:my_label}
\end{figure}

    \begin{figure}[H]
    \begin{verbatim}
stap_glm
 family:       gaussian [identity]
 formula:      y ~ sex + sap(Fast_Food)
 observations: 950
 Intercept:  TRUE
 fixed predictors:   1
 spatial predictors:  1
 temporal predictors:  0
 spatial-temporal predictors:  0
------
                        Median MAD_SD
(Intercept)             22.4    0.5  
sexF                    -0.7    0.1  
Fast_Food                1.2    0.1  
Fast_Food_spatial_scale  0.5    0.1  

Auxiliary parameter(s):
      Median MAD_SD
sigma 2.2    0.1   

Sample avg. posterior predictive distribution of y:
         Median MAD_SD
mean_PPD 24.4    0.1  

------
* For help interpreting the printed output see ?print.stapreg
* For info on the priors used see ?prior_summary.stapreg
\end{verbatim}
\caption{\texttt{stap\_glm} Model printout}
\label{fig:stapglmprintout}
\end{figure}

\subsection*{Simulation II: Correlated Outcomes}

Our second simulation showcases the typical STAP data structure in a longitudinal setting. Consider a model similar to the previous except we now have multiple measurements on each subject, in addition to the time each subject spent at their respective locations.
\begin{align*}
BMI_{i,j|b_i} &= \alpha + Z_{i,j}^{T}\delta + X_{i,j}(\theta^{s}=.8,\theta^t=18)\beta + b_i + \epsilon_{ij}  \quad j = 1,2\\
 b_i &\sim N(0,1.5)\\
 \epsilon_{ij} &\sim N(0,2)
\end{align*}
Prior to model fitting, the data collected for subjects and BEFs can be quite complex, as subjects move across space and time and businesses will open and close rendering a dynamic covariate space. Structurally, however, the data submitted to \texttt{rstap} will be similar to the uncorrelated spatial setting demonstrated in Simulation I, with the addition of at least one new column for a ``group ID'' so that distances or times for specific BEFs are associated with the subject at the appropriate measurement or group level. Note that multiple groups may be included in the time or distance data frame as coded ID column variables and, consequently, multiple IDs passed to the \texttt{stap\_glmer} ``group\_ID'' argument. See Tables \ref{tab:distdata2} and \ref{tab:timedata} for how the two data frames supplied to \texttt{distance\_data} and \texttt{time\_data}, respectively, are structured in a possible setting where a subject moves locations between two consecutive measurements. \\

\begin{table}[H]
 
           \captionsetup[subtable]{position = below}
          \centering
          \hspace{-2cm}
           \begin{subtable}{0.3\linewidth}
               \centering
               \footnotesize
               \begin{tabular}{c|c|c|c|c}
                s\_ID & m\_ID & bef\_ID & bef\_name    & Distance \\ \hline
                1           & 1           & 1       & CoffeeShop   & 0.351    \\
                1           & 1           & 2       & CoffeeShop   & 0.891    \\
                1           & 2           & 1       & CoffeeShop   & 0.413    \\
                1           & 2           & 1       & CoffeeShop   & 1.343
                \end{tabular} 
                \caption{Example distance data structure}
                \label{tab:distdata2}
                \end{subtable}
                \hspace*{3.5 cm}
                \begin{subtable}{.3 \linewidth}
                \centering
                \footnotesize
               \begin{tabular}{c|c|c|c|c}
                s\_ID & m\_ID & bef\_ID & bef\_name    & Time \\ \hline
                1           & 1           & 1       & CoffeeShop   & 3.43    \\
                1           & 1           & 2       & CoffeeShop   & 2.891    \\
                1           & 2           & 1       & CoffeeShop   & 0.513    \\
                1           & 2           & 2  & CoffeeShop   & 0.513
                \end{tabular}
                \caption{Example time data structure}
                \label{tab:timedata}
            \end{subtable}
            \caption{Example Longitudinal Built-Environment data structures as might be used in \texttt{rstap}}

\end{table}

\begin{figure}[H]
    \centering
    \includegraphics[width = .45 \textwidth]{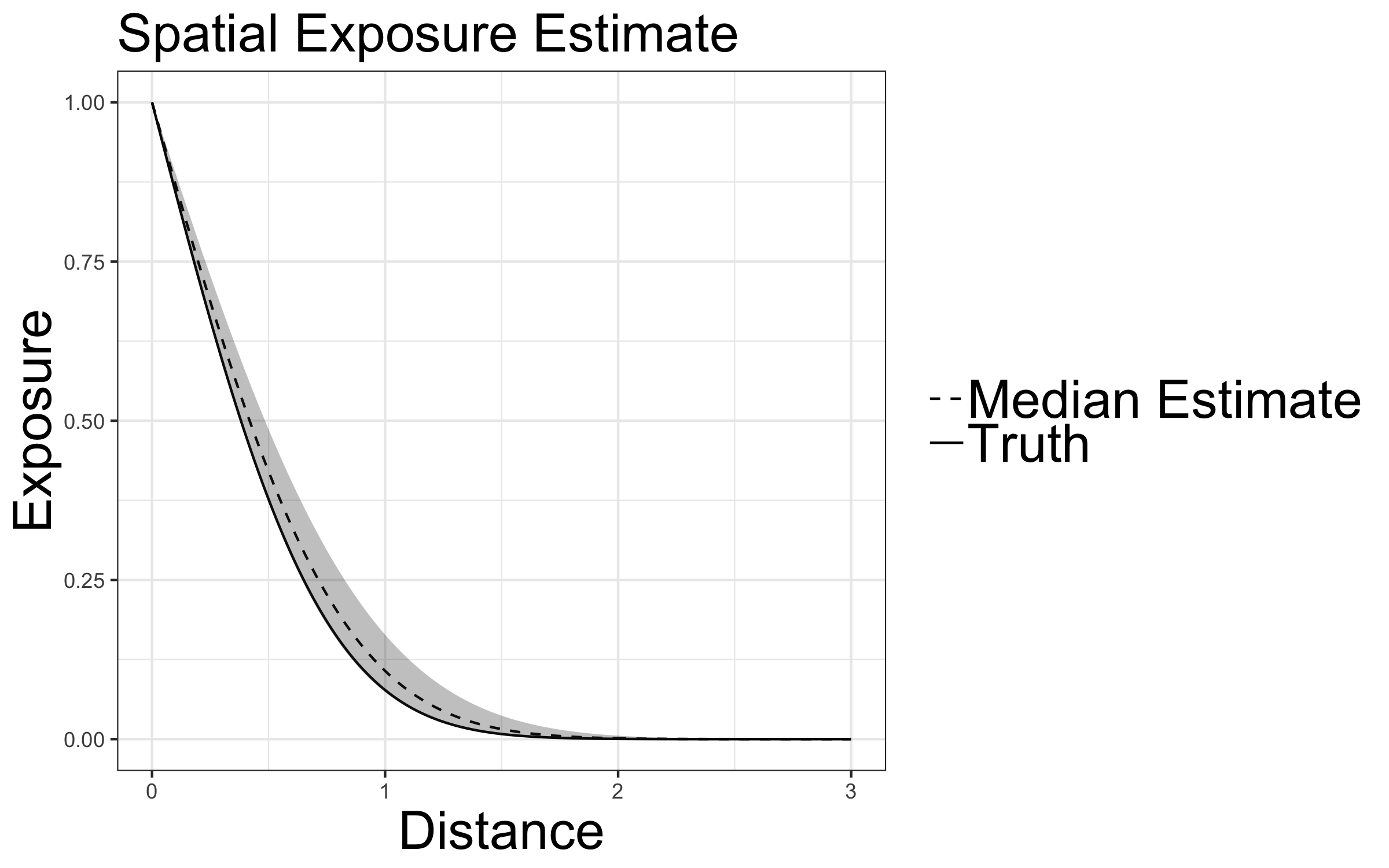}
    \includegraphics[width = .45 \textwidth]{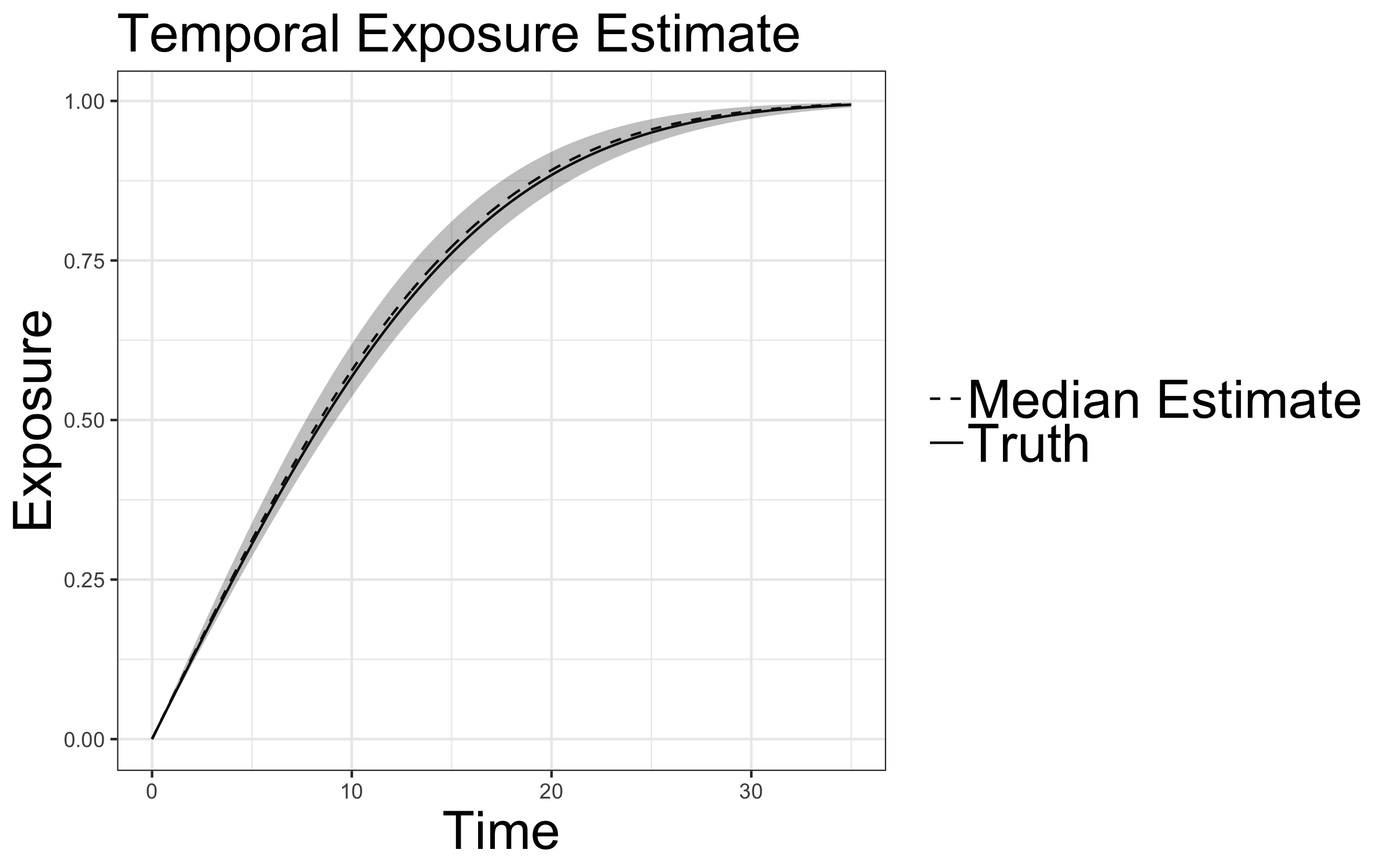}
    \caption{Longitudinal Simulation Spatial-Temporal exposure estimates with shaded area corresponding to 95\% of posterior. }
    \label{fig:longsimscales}
\end{figure}

\indent Simulating data under the model above with the same spatial configuration as the first and setting $\beta = 1$, we fit the model drawing two thousand samples on four independent chains. The first thousand samples from each chain are used for warm-up resulting in four thousand total posterior samples.  We use similar priors as before and place priors on the subject specific variance parameter in accordance with \texttt{rstanarm}'s recommendations for hierarchical covariance matrices \cite{rstanarm,gelman2013bayesian}.  The model output printed from \texttt{rstap} can be found below along with a visualization of both the spatial temporal exposure estimates in Figure \ref{fig:longsimscales}. A full workup for how this kind of data may be generated and compiled can be found at the package's website along with other, more complicated, simulated spatial structures and weight functions  (https://biostatistics4socialimpact/rstap).

\begin{figure}[H]
\begin{verbatim}
stap_glmer
 family:       gaussian [identity]
 formula:      y ~ sex + stap(Coffee_Shop) + (1 | subj_ID)
 observations: 658
 Intercept:  TRUE
 fixed predictors:   1
 spatial predictors:  0
 temporal predictors:  0
 spatial-temporal predictors:  1
------
                           Median MAD_SD
(Intercept)                21.2    0.4  
sex                         1.3    0.2  
Coffee_Shop                 0.9    0.1  
Coffee_Shop_spatial_scale   0.9    0.1  
Coffee_Shop_temporal_scale 17.4    0.9  

Auxiliary parameter(s):
      Median MAD_SD
sigma 2.4    0.1   

Error terms:
 Groups   Name        Std.Dev.
 subj_ID  (Intercept) 1.6     
 Residual             2.4     
Num.levels: subj_ID 350 

Sample avg. posterior predictive distribution of y:
         Median MAD_SD
mean_PPD 29.9    0.2  

------
* For help interpreting the printed output see ?print.stapreg
* For info on the priors used see ?prior_summary.stapreg
\end{verbatim}
\caption{Model output for Simulation II}
\end{figure}

    \subsection*{California FitnessGram and NETS data}
    We use FitnessGram data for 5th grade students in California during the 2010-2011 Academic year to examine associations between availability of fast food restaurants near their school and children's obesity. Publicly available data files from the California Department of Education (CDE) contain information on students' obesity status, grouped at the school level, or within sex and race/ethnicity groups within the school\cite{CDE}. We also obtained school-level covariates from the CDE, namely, charter status, percent of children eligible for free or reduce priced meals, and majority race/ethnicity of the school. We joined this data set to the Census track characteristics where the school was located, to include school-neigbhorhood characteristics: median household income and level of urbanization. Level of urbanicity around the school was coded as Urban, Suburban and Rural. For this analysis, we subset the data to only examine schools within urban areas.
    
    Participating in the FitnessGram test is required by the State of California, and as such, informed consent is not required. Moreover, since all personal identifiers are removed by the CDE prior to making the school-level aggregated data publicly available to researchers, this research is considered exempt from ethics review by the institutional review boards of The University of Michigan.
    
    The locations of California FFRs that were open during this time period were obtained from a commercial source\cite{walls2013national}. Distances were calculated between each school and businesses within ten miles categorized as ``Fast-Food Chain'' or ``Fast-Food Non-Chain''. These two business classes were combined to form one class, ``Fast Food Restaurants`` (FFR), to be included in the model. The ten mile inclusion distance was chosen as a conservative upper bound as previous associations estimated with a DLM decline to zero at far shorter distances \cite{baek2016distributed}.
    
    We fit the following model in order to estimate the odds of obesity for a fifth grade child  as function of their Fast Food environment, after adjusting for the aforementioned confounders:
    
    \begin{align*}
    \text{logit}(\mu_i) &= \alpha + Z_{i}^{T}\delta + X_{i,FFR}(\theta_s)\beta  \\
    \alpha &\sim N(0,3)  \quad \quad \delta \sim N(0,2) \\
    \beta &\sim N(0,2) \quad \quad \log(\theta^s) \sim N(1,1)
    \end{align*}
     \noindent In this example we chose priors for the prior intercept and regression coefficients according to Gelman et al's principles for weakly informative priors in a logistic regression setting\cite{gelman2013bayesian,gelman2015stan}. We set the prior for the STAP regression coefficient similarly using previous research \cite{baek2016distributed} for further justification. Corresponding to our inclusion distance of 10 miles, a prior distribution was set on the spatial scale such that the median exposure was 0.01 - effectively negligible - after 5 miles but whose tail was long enough that a true higher scale would not be as severely penalized as it would with, say, a normal prior. Note that we do not use uninformative priors in this setting as it not only makes sampling times prohibitive, but corresponds to placing substantial probability mass on many scales that would set equivalent exposures to BEFs both close and extremely far away from schools.  We code this prior information and model in \texttt{rstap}, with the function call seen in Figure \ref{fig:glmbinom}.
    \begin{figure}[H]
    \begin{verbatim}
        stap_glm(formula = cbind(Num_Obese,Num_NotObese) ~ Charter_I + 
                            MedianIncome_centered + Majority_Race + Percent_Educated +
                            frpm +  sap(FFR),
                 family= binomial(link='logit'),
                 subject_data = subj_df,
                 distance_data = dsts_df,
                 subject_ID = "school_ID",
                 max_distance = 10, 
                 prior = normal(location = 0, scale = 2, autoscale = F),
                 prior_intercept = normal(location =0, scale = 3, autoscale = F),
                 prior_stap = normal(location = 0, scale = 2),
                 prior_theta = log_normal(location = 1,scale = 1), 
                 chains = 4, cores = 4, iter = 2E3)
    \end{verbatim}
    \caption{Syntax for fitting binomial model - CA data}
    \label{fig:glmbinom}
    \end{figure}
    A visualization for the differing model's estimates of the spatial scale can be found in Figure \ref{fig:CA_scales} along with the initial prior distribution and FFR effect. \\

    \begin{figure}[H]
        \centering
        \begin{subfigure}{0.48\textwidth}
\centering
        \includegraphics[width =  \textwidth]{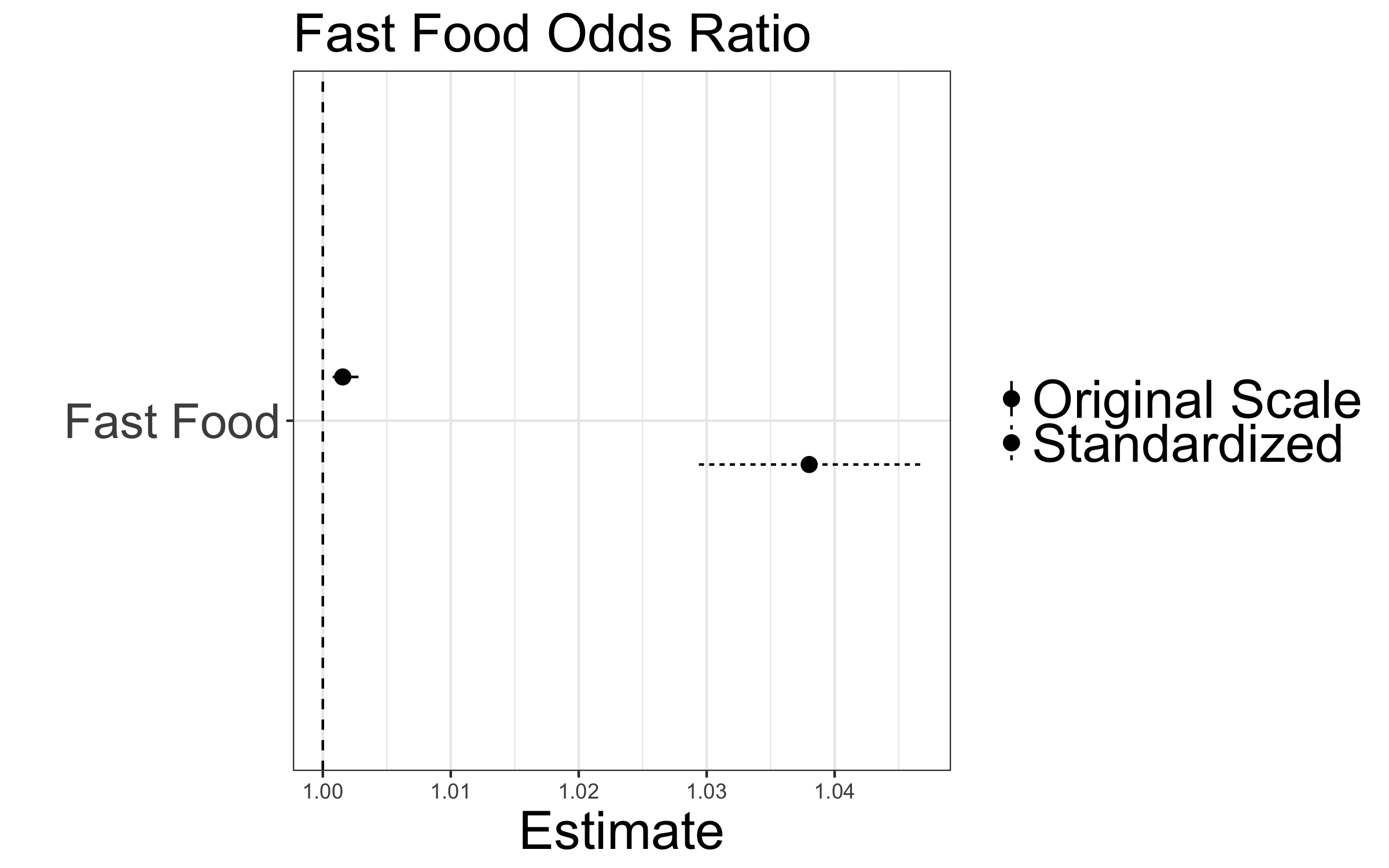}
    \caption{Fast Food Effect Estimate}
    \label{fig:FFE}
\end{subfigure}
\begin{subfigure}{0.48\textwidth}
\centering
        \includegraphics[width =  \textwidth]{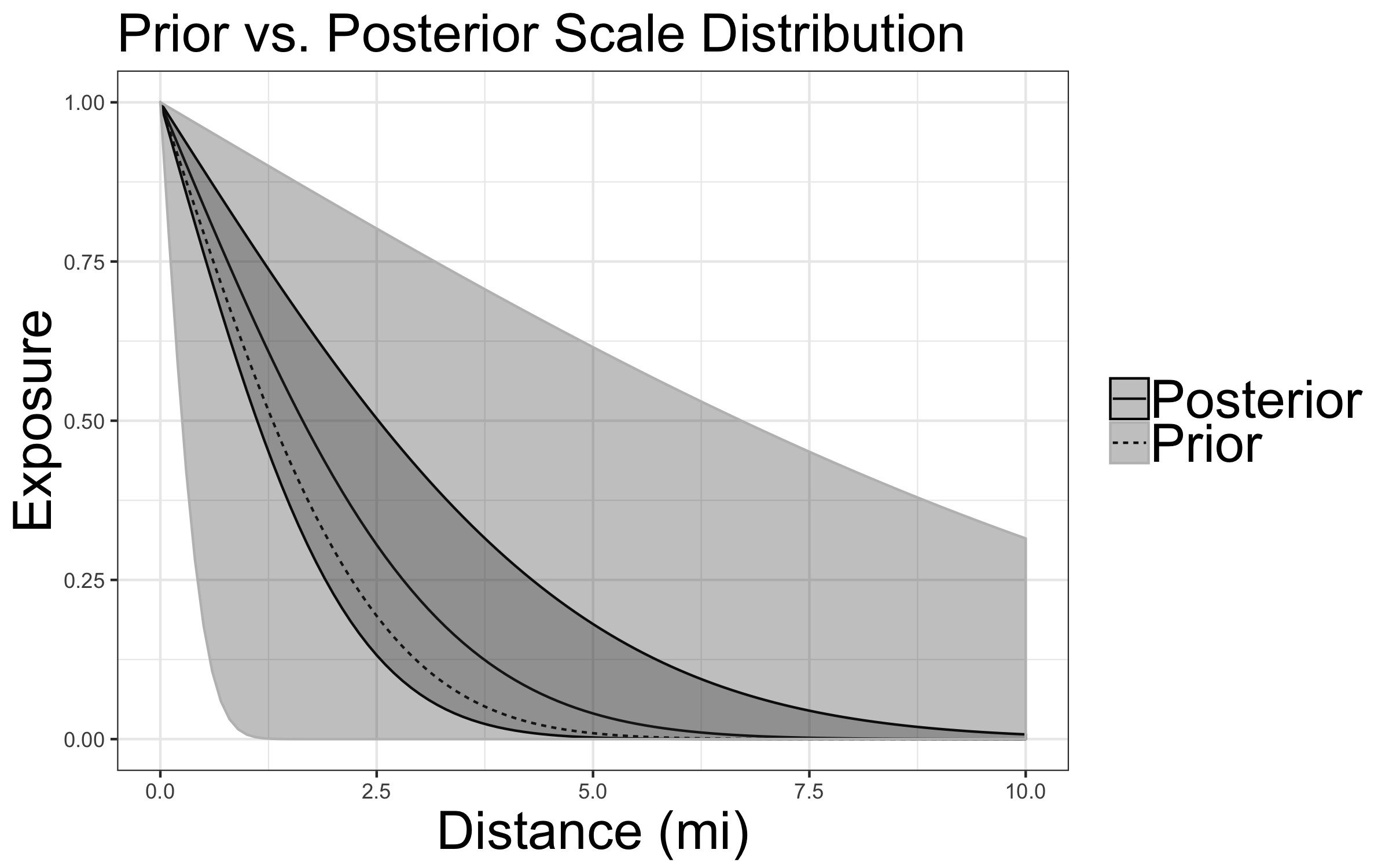}
    \caption{Spatial Exposure Prior and Posterior Estimate}
    \label{fig:spatialscale}
\end{subfigure}
        \caption{Fast Food effect and exposure estimates in Urban CA, alongside spatial prior. Shaded area corresponds to 95\% of posterior}
        \label{fig:CA_scales}
    \end{figure}
    
\begin{table}[H]
\centering 
\caption{Distance at which Exposure Effect is negligible (mi)}
\begin{tabular}{rrrr}
  \hline
 & 2.5\% & 50\% & 97.5\% \\ 
  \hline
  Urban & 4.27 & 6.28 & 9.62 \\ 
   \hline
\end{tabular}
\label{tab:exptermest}
\end{table}
    
    \indent We found that higher exposure to FFRs near schools is associated with a higher odds of obesity after adjusting for confounders.  In urban areas, one FFR placed zero miles from a school is associated with an odds ratio of 1.001558 (95 \% CI 1.000471 to 1.002783). The small magnitude of this estimate reflects the saturation of FFRs in urban environments, i.e. one more FFR has negligible association with obesity. To emphasize this fact, the standard deviation of the $X(\theta^s)$ covariate in urban areas is 23.8 (12.59 to 51.8), implying that if 24 stores were added at zero miles from a school - one standardized effect - the odds ratio for obesity would be roughly 4\% as seen in Figure \ref{fig:CA_scales} (a). \\
    \indent Moreover, using our spatial scale estimates we can see that the effect of one additional restaurant is effectively negligible after, roughly,  six miles. We calculate this precisely for Table \ref{tab:exptermest}, using the \texttt{stap\_termination} function in \texttt{rstap}. This function requires a specification of what exposure to consider ``effectively negligible". For Table \ref{tab:exptermest} we use a value of 0.01  (code in Appendix, Figure \ref{fig:stapterm}). That is, we find the value $d$ for which $\mathcal{K}(\frac{d}{\hat{\theta}^s}) = 0.01$ for the corresponding $\hat{\theta^s}$ estimates.

\subsection*{FitnessGram data - Correlated outcomes}
\begin{figure}[H]
    \centering
    \begin{subfigure}{.45 \textwidth}
    \centering
    \includegraphics[width = .95\textwidth]{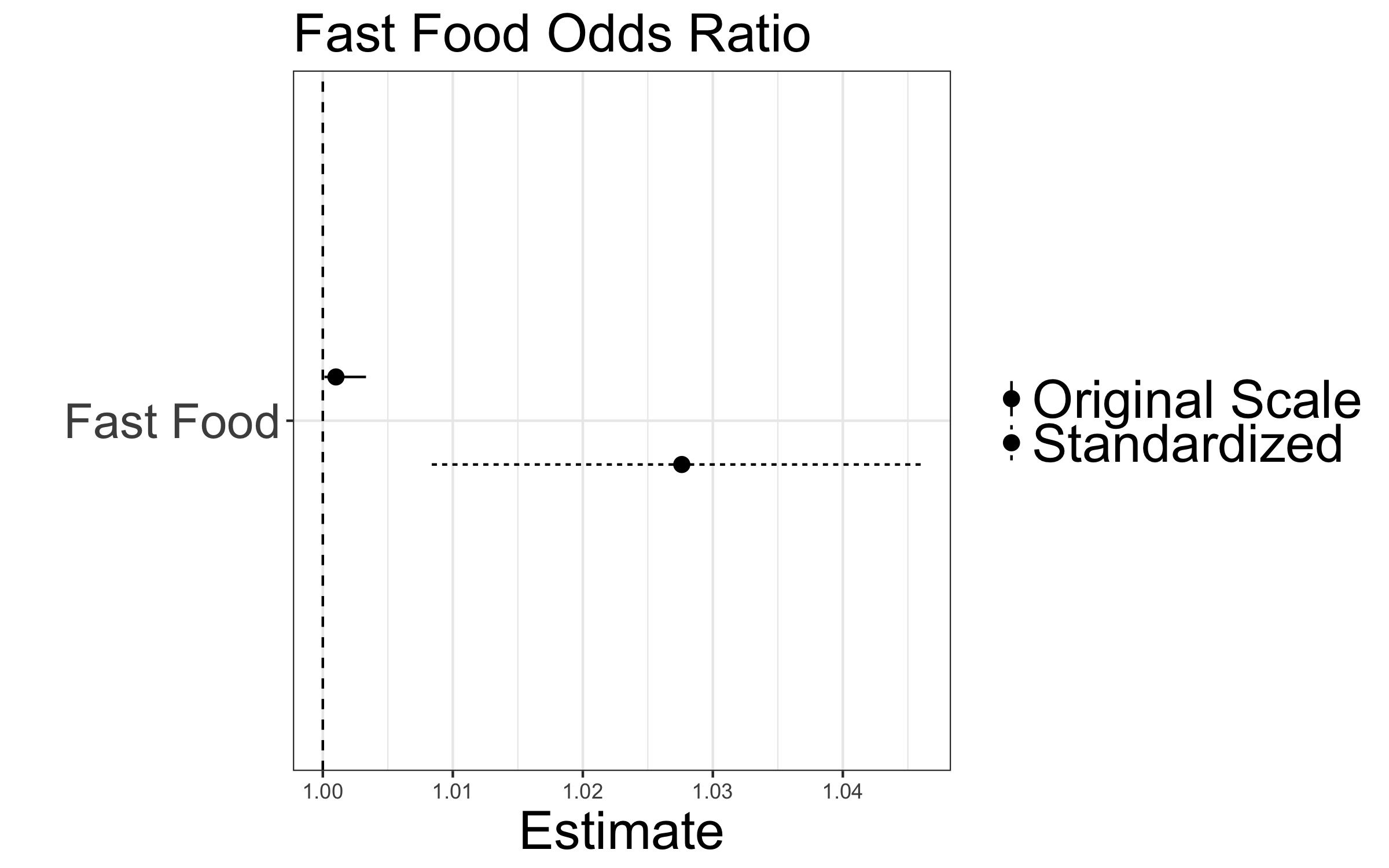}
    \caption{Fast Food Effect}
    \end{subfigure}
    \begin{subfigure}{.45 \textwidth}
    \centering
        \includegraphics[width =  .95\textwidth]{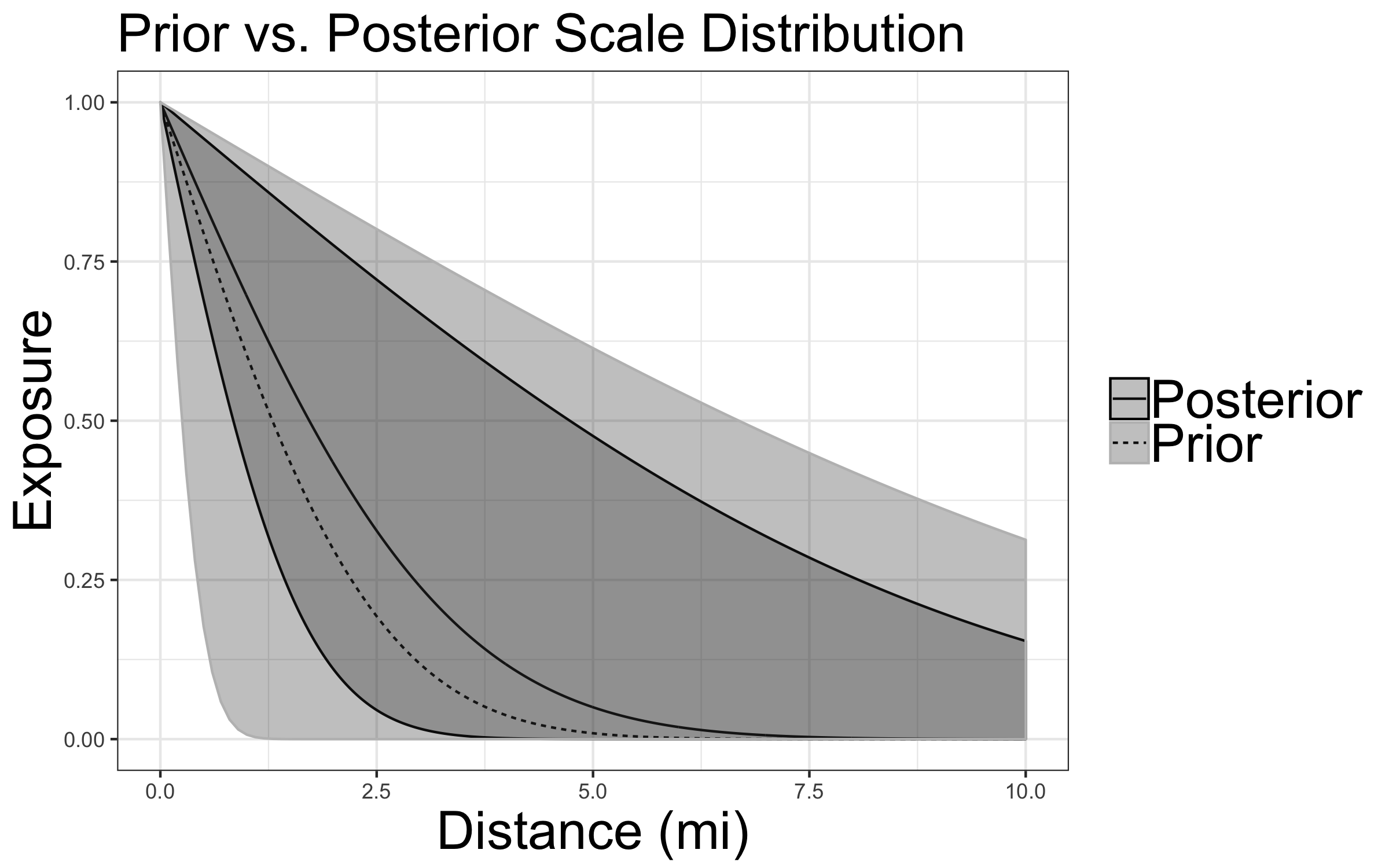}
        \caption{Urban Scale}
    \end{subfigure}

        \caption{Fast Food Urban Scale and Exposure Main Effect. Shaded area corresponds to 95\% of posterior }
        \label{fig:CA_FF_corr}
    \end{figure}

To demonstrate how STAPs can be used with correlated binomial outcomes, we use the number of obese boys and girls at each school in the same setting as before. In terms of data structure, since the regression design matrix must be broken out into two groups, the distance data submitted to \texttt{rstap} must also be broken out, or in this case, copied into two groups to reflect the corresponding spatial temporal exposure for the observation at the appropriate group level. The data are copied in this setting because the BEF exposure is the same for the school, regardless of whether the boy or girl outcome is of interest. In this case, the modification is reflected in the \texttt{Gender\_CAT} ID key and covariate shown in the \texttt{stap\_glmer} function below. This function fits the model with a random intercept and similar priors placed on $\tau$ as in our longitudinal simulation and $\bm{\delta}$ as in the previous example. Additionally, the same visualizations are provided for this new model in Figure \ref{fig:CA_FF_corr}.
    
    \begin{align*}
    \text{logit}(\mu_{i,j|b_i} ) &= \alpha + Z_{i,j}^{T}\delta + X_{i,FFR}^{T}(\theta_s)\beta + b_i  \\
    j &\text{ denotes an outcome for girls or boys at the } i \text{th} \text{ school}\\
    b_i &\sim N(0,\tau) \tag{priors in function call}
    \end{align*}
    
    \begin{figure}[H]
    \begin{verbatim}
    stap_glmer(formula = cbind(NoStud5c,NoStud_NObese) ~ Charter_I + 
                            MedianIncome_centered + Majority_Race + Gender_CAT + 
                            Percent_Educated +  frpm + sap(FFR) + (1|school_ID),
                 family= binomial(link="logit"),
                 subject_data = subj_df,
                 distance_data = dsts_df,
                 subject_ID = "school_ID",
                 group_ID = "Gender_CAT"
                 max_distance = 10, 
                 prior = normal(location = 0, scale = 2),
                 prior_intercept = normal(location = 0, scale = 3),
                 prior_stap = normal(location = 0, scale = 2),
                 prior_theta = log_normal(location = 1,scale = 1), 
                 prior_covariance = decov(regularization = 1,
                                          concentration = 1,
                                          shape = 1, scale = 1)),
                 chains = 4, cores = 4, iter = 2E3)
    \end{verbatim}
    \caption{Model syntax for fitting California Data with correlated outcomes}
    \end{figure}

    Examining this model's estimates we can see that a similar median exposure scale is attained, with a much wider uncertainty. This is likely due to the increased number of parameters being estimated in this new model formulation, in addition to the increased uncertainty imposed by the introduction of an additional latent parameter into the model.
    
    \section{Discussion}
    
	 The present paper is meant to provide a general overview of the R package \texttt{rstap} implementing STAPs using the probabilistic programming language \texttt{Stan} for full Bayesian inference. This article provides a starting point for the reader interested in understanding how to fit basic STAP models.\par
	  The current extension planned for the next release of \texttt{rstap} is a custom HMC sampler with symbolic derivatives of the log probability hard coded for typical STAP model configurations that, not requiring automatic differentiation, will sample from the posterior much more quickly. This will be advantageous for modeling scenarios in which there are a large number of STAPS to be fit and/or a large number of built environment features associated with each subject. Other extensions could include providing functions for modeling stap-interactions in a setting where it is believed a scale may differ among levels of some categorical, e.g. gender, covariate.
	  \section{Acknowledgements}
	  The authors would like to acknowledge funding sources RO1 HL131610 and RO1 HL136718 in support of this work.

\bibliographystyle{unsrt}
\bibliography{rstap}

 \newpage
	\section*{Appendix}

\subsection*{Spatial Temporal Scale Upper Bound Construction}
    For weighting function $\mathcal{K}_j(d)$ associated with each STAP $j=1,...,J$ we find $\theta^{up}$, the upper bound used in estimating $\theta_{j}^{s}$, $j = 1,...,J$. Here, $d_j^*$ is the distance specific to the specified $\mathcal{K}$ for class $j$ at which the evaluation is equal to 0.975. Since different staps may have different weight functions, and consequently, different $d_j^*$'s taking the minimum across these distances results in a unified upper bound for all $\theta$, which is convenient computationally.
    \begin{figure}[H]
        \begin{align*}
        \mathcal{K}_j(d^{*}_j) &= 0.975 \implies \theta^{up} := \frac{\text{max distance}}{\min_j(d^{*}_j)} \tag{4}
        \end{align*}
        \caption{Spatial-Temporal Scale Upper Bound Construction}
        \label{fig:theta_up}
    \end{figure}

\begin{figure}[H]
    \begin{verbatim}
    reps <- posterior_predict(fit)
    ## posterior predictive plot
    pp_plot <- bayesplot::ppc_dens_overlay(y = fit$y, yrep = reps) 
    nuts_diagnostics <- bayesplot::nuts_params(fit$stapfit)
    ## NUTS energy diagnostic plot
    nuts_plot <- bayesplot::mcmc_nuts_energy(nuts_diagnostics) 
    \end{verbatim}
    \caption{Code for producing posterior predictive and NUTS diagnostic plots}
    \label{fig:ppcs}
\end{figure}

\begin{figure}[H]
\begin{verbatim}
    s1 <- stap_termination(fit_rural, exposure_limit = 0.01, prob = .95, max_value = 15)
    s2 <- stap_termination(fit_suburb, exposure_limit = 0.01, prob = 0.95, max_value = 15)
    s3 <- stap_termination(fit_urban, exposure_limit = 0.01, prob = 0.95, max_value = 15)
    tbl <- rbind(s1,s2,s3)
    rownames(tbl) <- c("Rural","Suburban","Urban")
    colnames(tbl) <- c("2.5%","50%","97.5%")
    xtable::xtable(tbl,caption="Spatial Exposure Termination Estimates")
\end{verbatim}
\caption{Code for producing \texttt{stap\_termination} values}
\label{fig:stapterm}
\end{figure}

\begin{figure}[H]
\begin{verbatim}
R> summary(fit,waic=T)
Model Info:

 function:     stap_glm
 family:       gaussian [identity]
 formula:      y ~ sex + sap(Fast_Food)
 priors:       see help('prior_summary')
 sample:       4000 (posterior sample size)
 observations: 950
 Spatial Predictors:   1
 WAIC: 4220

Estimates:
                          mean    sd      2.5%    25%     50%     75%     97.5%
(Intercept)                22.3     0.5    21.3    22.0    22.4    22.7    23.1
sexF                       -0.7     0.1    -1.0    -0.8    -0.7    -0.6    -0.4
Fast_Food                   1.2     0.1     1.0     1.1     1.2     1.2     1.4
Fast_Food_spatial_scale     0.5     0.1     0.4     0.5     0.5     0.6     0.6
sigma                       2.2     0.1     2.1     2.2     2.2     2.3     2.3
mean_PPD                   24.4     0.1    24.2    24.3    24.4    24.4    24.6
log-posterior           -2117.4     1.6 -2121.2 -2118.2 -2117.0 -2116.2 -2115.3

Diagnostics:
                        mcse Rhat n_eff
(Intercept)             0.0  1.0  4000 
sexF                    0.0  1.0  4000 
Fast_Food               0.0  1.0  4000 
Fast_Food_spatial_scale 0.0  1.0  4000 
sigma                   0.0  1.0  4000 
mean_PPD                0.0  1.0  4000 
log-posterior           0.0  1.0  1714 

For each parameter, mcse is Monte Carlo standard error, n_eff is a crude measure
of effective sample size, and Rhat is the potential scale reduction factor
on split chains (at convergence Rhat=1).
\end{verbatim}
\caption{Summary print out - simulated model section 4.4.1}
\label{fig:sumex}
\end{figure}

\end{document}